\input harvmac\skip0=\baselineskip
\input epsf

\newcount\figno
\figno=0
\def\fig#1#2#3{
\par\begingroup\parindent=0pt\leftskip=1cm\rightskip=1cm\parindent=0pt
\baselineskip=11pt \global\advance\figno by 1 \midinsert
\epsfxsize=#3 \centerline{\epsfbox{#2}} \vskip 12pt {\bf Fig.\
\the\figno: } #1\par
\endinsert\endgroup\par
}
\def\figlabel#1{\xdef#1{\the\figno}}
\def\encadremath#1{\vbox{\hrule\hbox{\vrule\kern8pt\vbox{\kern8pt
\hbox{$\displaystyle #1$}\kern8pt} \kern8pt\vrule}\hrule}}

\def\p{\partial}


\lref\orv{
A.~Okounkov, N.~Reshetikhin and C.~Vafa,
``Quantum Calabi-Yau and classical crystals,''
arXiv:hep-th/0309208.
} \lref\dff{ V.~de Alfaro, S.~Fubini and G.~Furlan, `Conformal
Invariance In Quantum Mechanics,'' Nuovo Cim.\ A {\bf 34}, 569
(1976). }
 \lref\gssy{D.~Gaiotto, A.~Simons, A.~Strominger and X.~Yin,
``D0-branes in Black Hole Attractors,'' arXiv:hep-th/0412179.}
 \lref\gt{
G.~W.~Gibbons and P.~K.~Townsend, ``Black holes and Calogero
models,'' Phys.\ Lett.\ B {\bf 454}, 187 (1999)
[arXiv:hep-th/9812034].
}
\lref\ssty{
A.~Simons, A.~Strominger, D.~M.~Thompson and X.~Yin,
``Supersymmetric branes in AdS(2) x S**2 x CY(3),''
arXiv:hep-th/0406121.
}

\lref\rgcv{
R.~Gopakumar and C.~Vafa,
``M-theory and topological strings. I,''
arXiv:hep-th/9809187.
}
\lref\aes{
A.~Strominger,
``Macroscopic Entropy of $N=2$ Extremal Black Holes,''
Phys.\ Lett.\ B {\bf 383}, 39 (1996)
[arXiv:hep-th/9602111].
}
\lref\MohauptMJ{
T.~Mohaupt,
``Black hole entropy, special geometry and strings,''
Fortsch.\ Phys.\  {\bf 49}, 3 (2001)
[arXiv:hep-th/0007195].
} \lref\shmak{ M.~Shmakova, `Calabi-Yau black holes,'' Phys.\
Rev.\ D {\bf 56}, 540 (1997) [arXiv:hep-th/9612076].
}
\lref\iceland{ R.~Britto-Pacumio, J.~Michelson, A.~Strominger
and A.~Volovich, ``Lectures on superconformal quantum mechanics
and multi-black hole  moduli spaces,'' arXiv:hep-th/9911066.
}

\lref\mss{
A.~Maloney, M.~Spradlin and A.~Strominger,
``Superconformal multi-black hole moduli spaces in four dimensions,''
JHEP {\bf 0204}, 003 (2002)
[arXiv:hep-th/9911001].
}

\lref\msas{M.~Spradlin and A.~Strominger, ``Vacuum states for
AdS(2) black holes,'' JHEP {\bf 9911}, 021 (1999)
[arXiv:hep-th/9904143].} \lref\juanbh{J.~M.~Maldacena, ``N = 2
extremal black holes and intersecting branes,'' Phys.\ Lett.\ B
{\bf 403}, 20 (1997) [arXiv:hep-th/9611163].} \lref\jmjmas{\
J.~M.~Maldacena, J.~Michelson and A.~Strominger, ``Anti-de Sitter
fragmentation,'' JHEP {\bf 9902}, 011 (1999)
[arXiv:hep-th/9812073]. } \lref\fvijay{V.~Balasubramanian and
F.~Larsen, ``On D-Branes and Black Holes in Four Dimensions,''
Nucl.\ Phys.\ B {\bf 478}, 199 (1996) [arXiv:hep-th/9604189].}
\lref\mstwo{
J.~Michelson and A.~Strominger,
``Superconformal multi-black hole quantum mechanics,''
JHEP {\bf 9909}, 005 (1999)
[arXiv:hep-th/9908044].
}

\lref\msone{
J.~Michelson and A.~Strominger,
``The geometry of (super)conformal quantum mechanics,''
Commun.\ Math.\ Phys.\  {\bf 213}, 1 (2000)
[arXiv:hep-th/9907191].
}

\lref\fks{
S.~Ferrara, R.~Kallosh and A.~Strominger,
``N=2 extremal black holes,''
Phys.\ Rev.\ D {\bf 52}, 5412 (1995)
[arXiv:hep-th/9508072].
} \lref\myers{} \lref\bsv{ R.~Britto-Pacumio, A.~Strominger and
A.~Volovich, ``Two-black-hole bound states,'' JHEP {\bf 0103}, 050
(2001) [arXiv:hep-th/0004017].
}
\lref\bmss{
R.~Britto-Pacumio, A.~Maloney, M.~Stern and A.~Strominger,
``Spinning bound states of two and three black holes,''
JHEP {\bf 0111}, 054 (2001)
[arXiv:hep-th/0106099].
}

\lref\VafaGR{
C.~Vafa,
``Black holes and Calabi-Yau threefolds,''
Adv.\ Theor.\ Math.\ Phys.\  {\bf 2}, 207 (1998)
[arXiv:hep-th/9711067].
}
\lref\myers{
R.~C.~Myers,
``Dielectric-branes,''
JHEP {\bf 9912}, 022 (1999)
[arXiv:hep-th/9910053].
}

\lref\myersRV{
R.~C.~Myers,
``Nonabelian Phenomena on D-branes'',
Class.Quant.Grav.20, S347-S372 (2003)
[arXiv:hep-th/0303072].
}

\lref\osv{ H.~Ooguri, A.~Strominger and C.~Vafa, ``Black hole
attractors and the topological string,'' arXiv:hep-th/0405146.
} \lref\spinor{T. Mohaupt, ``Black Hole Entropy, Special Geometry
and Strings", hep-th/0007195.}

\lref\bpsb{ M. Marino, R. Minasian, G. Moore and A. Strominger,
``Nonlinear Instantons from Supersymmetric $p$-Branes",
hep-th/9911206. }

\lref\msw{ J. Maldacena, A. Strominger and E. Witten, ``Black Hole
Entropy in M-Theory", hep-th/9711053. } \lref\gssy{ D. Gaiotto,
A.~Simons, A.~Strominger and X.~Yin, "D0-branes in black hole
attractors", arXiv:hep-th/0412179.} \lref\SimonsNM{ A.~Simons,
A.~Strominger, D.~M.~Thompson and X.~Yin, ``Supersymmetric branes in
AdS(2) x S**2 x CY(3),'' arXiv:hep-th/0406121.
}
\lref\suss{
B.~Freivogel, L.~Susskind and N.~Toumbas,
``A two fluid description of the quantum Hall soliton,''
arXiv:hep-th/0108076.
}

\lref\claus{ P.~Claus, M.~Derix, R.~Kallosh, J.~Kumar,
P.~K.~Townsend and A.~Van Proeyen, ``Black holes and
superconformal mechanics,'' Phys.\ Rev.\ Lett.\  {\bf 81}, 4553
(1998) [arXiv:hep-th/9804177].} \lref\juan{ J.~M.~Maldacena, `The
large N limit of superconformal field theories and supergravity,''
Adv.\ Theor.\ Math.\ Phys.\ {\bf 2}, 231 (1998) [Int.\ J.\ Theor.\
Phys.\  {\bf 38}, 1113 (1999)] [arXiv:hep-th/9711200]. }

\lref\gqkahler{ N. Reshetikhin and L. Takhtajan, ``Deformation
Quantization of K\"ahler Manifolds'', math.QA/9907171. }

\Title{\vbox{\baselineskip12pt\hbox{hep-th/0412322} }}{
Superconformal Black Hole Quantum Mechanics}

\centerline{Davide Gaiotto,~ Andrew Strominger and Xi Yin }
\smallskip
\centerline{Jefferson Physical Laboratory, Harvard University,
Cambridge, MA 02138} \vskip .6in \centerline{\bf Abstract} {In
recent work, the superconformal quantum mechanics describing D0
branes in the $AdS_2\times S^2\times CY_3$ attractor geometry of a
Calabi-Yau black hole with D4 brane charges $p^A$ has been
constructed and found to contain a large degeneracy of chiral
primary bound states. In this paper it is shown that the asymptotic
growth of chiral primaries for $N$ D0 branes exactly matches the
Bekenstein-Hawking area law for a black hole with D4 brane charge
$p^A$ and D0 brane charge $N$. This large degeneracy arises from D0
branes in lowest Landau levels which tile the $CY_3\times S^2$
horizon. It is conjectured that such a multi-D0 brane $CFT_1$ is
holographically dual to IIA string theory on $AdS_2\times S^2\times
CY_3$.} \vskip .3in

\smallskip
\Date{}

Compactifications of type II string theory on a Calabi-Yau space
$CY_3$ contain extremal black hole solutions characterized by
electric and magnetic charges $(p^\Lambda, q_\Lambda)$. The near
horizon region is an $AdS_2\times S^2\times CY_3$ attractor
geometry\refs{\fks,\aes}. On general grounds \juan, one expects
the black hole attractor to be holographically dual to a
conformally invariant quantum mechanics. However, despite enormous
progress on $AdS/CFT$ in higher dimensions, an understanding of
$AdS_2/CFT_1$ has remained elusive.  To date no complete examples
are known.

 In this paper we will pull together several
observations and propose an $AdS_2/CFT_1$ correspondence, bringing
to partial fruition the program initiated in \refs{\msone \mstwo
\mss \bsv \bmss -\iceland}. In recent work \osv\ it has become clear
that Calabi-Yau black holes are most naturally described in terms of
fixed magnetic charges $p^\Lambda$ and a weighted ensemble of
electric charges $q_\Lambda$. This motivates the study of the
quantum mechanical partition function of (electric) D0-branes in the
background attractor geometry produced by (magnetic) D4 flux. The
multi-D0 quantum mechanics is acted on by the (super) isometries of
$AdS_2$ and is therefore automatically (super) conformally invariant
\refs{\claus,\gt}. The quantum mechanics exhibits a rich spectrum of
supersymmetric bound states, which can be described as D0-branes
which pop out and in of the black hole horizon \refs{\ssty,\gssy}.
Since they are not stationary with respect to asymptotic time, these
bound states preserve only near-horizon (but no asymptotic)
supersymmetries. Of particular interest are nonabelian $N$-D0
configurations corresponding to D2 branes which wrap the black hole
horizon and carry $N$ units of worldvolume magnetic flux. Such a
D2-D0 brane effectively sees a magnetic flux (proportional to the D4
flux) on $CY_3$ and thereby acquires a large degeneracy of lowest
Landau levels.  The corresponding chiral primary states are counted
and
 and found to exactly reproduce the leading order
area-entropy formula for a D0-D4 black hole.

Based on this agreement at large $N$ in the BPS sector,  we
propose that the superconformal multi-D0 quantum mechanics is the
holographic dual of string theory on the attractor geometry.
Although we do not do so herein, this proposal might be sharpened
and tested by examining subleading corrections to the entropy
formula. This duality might also be extended to non-BPS
excitations, taking due consideration of the subtleties in the
$AdS_2$ scaling limit\refs{\jmjmas,\msas}.

Of course, the problem of microscopically computing the entropy of
Calabi-Yau black holes has already been solved in the D0-D4 case
\refs{\msw,\VafaGR}(see also \refs{\juanbh, \fvijay}).  The new
features here are the appearance of a dual $CFT_1$ (potentially
relevant for non-BPS computations), and an interesting  physical
picture of the black hole microstates as D0 brane bound states
localized near the horizon. We also hope that the method will
generalize to Calabi-Yau black holes with a general set of
charges, for which no microscopic derivation of the entropy is
known.

To proceed, consider a black hole solution with D4 fluxes $p^A$,
$A=1,...b_2$ and D0 flux $q_0$.\foot{We will see later that the
background $q_0$ drops out of the formula for chiral primaries. At
this stage it is needed for a finite volume $CY_3$, but if we were
to include the one-loop correction to the prepotential the volume
may be finite in some cases even for $q_0=0$ \shmak.} According to
the attractor mechanism \refs{\fks,\aes}, the radius $R$ of the near
horizon $AdS_2\times S^2$ is equal to the graviphoton charge\foot{We
will be working in the convention $2\pi\sqrt{\alpha'}=1$, and the
ten-dimensional Newton's constant $G_N=g_s^2/32\pi^2$. In the
four-dimensional Planck units, we have $R=Q=(4Dq_0)^{1\over4}$. In
string units, $R={1\over 2}g_s \sqrt{D\alpha'/q_0}$.
 } $Q$ \eqn\fdt{R=Q={g_s\over 4\pi}\sqrt{ D \over
q_0},~~~~~~~~~~~~~~~~D\equiv D_{ABC}p^Ap^Bp^C,} with $6D_{ABC}$ the
triple intersection numbers on $CY_3$. The Kahler class of $CY_3$
 at the
horizon is \eqn\rts{J=\sqrt{q_0 \over D}{p^A} \omega_A,} where
$\omega_A$ is an integral basis for $H^2(CY_3)$. There are also RR
field strengths \eqn\rax{F^{(4)}=\omega_{S^2}p^A \omega_A,
~~~F^{(2)}=q_0 \omega_{AdS_2}} with the normalization
$\int_{S^2}\omega_{S^2}=1=\int_{S^2\times CY_3}*\omega_{AdS_2}$.

We wish to compute a partition function for $N$ D0 branes in the
black hole attractor geometry \fdt-\rax. Consider
 \eqn\pol{Z(q)\sim \Tr [{\cal O}q^N],} where
the trace is over the multi-D0 Hilbert space and $\cal O$ is an
appropriate operator insertion.
 In order to define and compute \pol\ we must choose a basis of
states. A natural choice is  eigenstates of the generator $H$ of
asymptotic black hole time, which corresponds to Poincar\'e time in
the near horizon $AdS_2$. However for this choice \pol\ is
generically  ill-defined already for a single D0. As discussed in
\refs{\bmss,\mstwo,\bsv,\iceland}, the problem arises from a
divergent spectrum of low-energy states with arbitrarily large
near-horizon redshifts.

To circumvent this problem, as in \refs{\gt, \bsv,\bmss} we note
that the D0-brane quantum mechanics has a (super)conformal structure
inherited from the (super)isometries of the $AdS_2$ spacetime in
which they move. This allows for the possibility \dff\ of using an
alternate basis in \pol : eigenstates of $L_0=H+K$, where $K$ is the
generator of special conformal transformations. $L_0$ generates
global (rather than Poincar\'e) time translations in $AdS_2$\gt. It
has a discrete spectrum and hence no problem with an IR divergent
continuum. $L_0$ eigenstates are not static with respect to
asymptotic black hole time, rather they pop out and in of the
horizon, as depicted in figure 1. Here we propose  that the black
hole ground states should be identified with the chiral primaries of
the near-horizon superconformal quantum mechanics.  We emphasize
that this identification is a proposal which should be tested and we
do not have a derivation of its validity. Further discussion and
motivation can be found in \bsv.

\fig{Penrose diagram for an extremal Calabi-Yau black hole. Each
point represents an $S^2\times CY_3$ except for the timelike
singularity on the right. The near-horizon $AdS_2$ region is
shaded, and the solid line within this region depicts the
trajectory of an $L_0$ eigenstate popping in and out of the
horizon.}{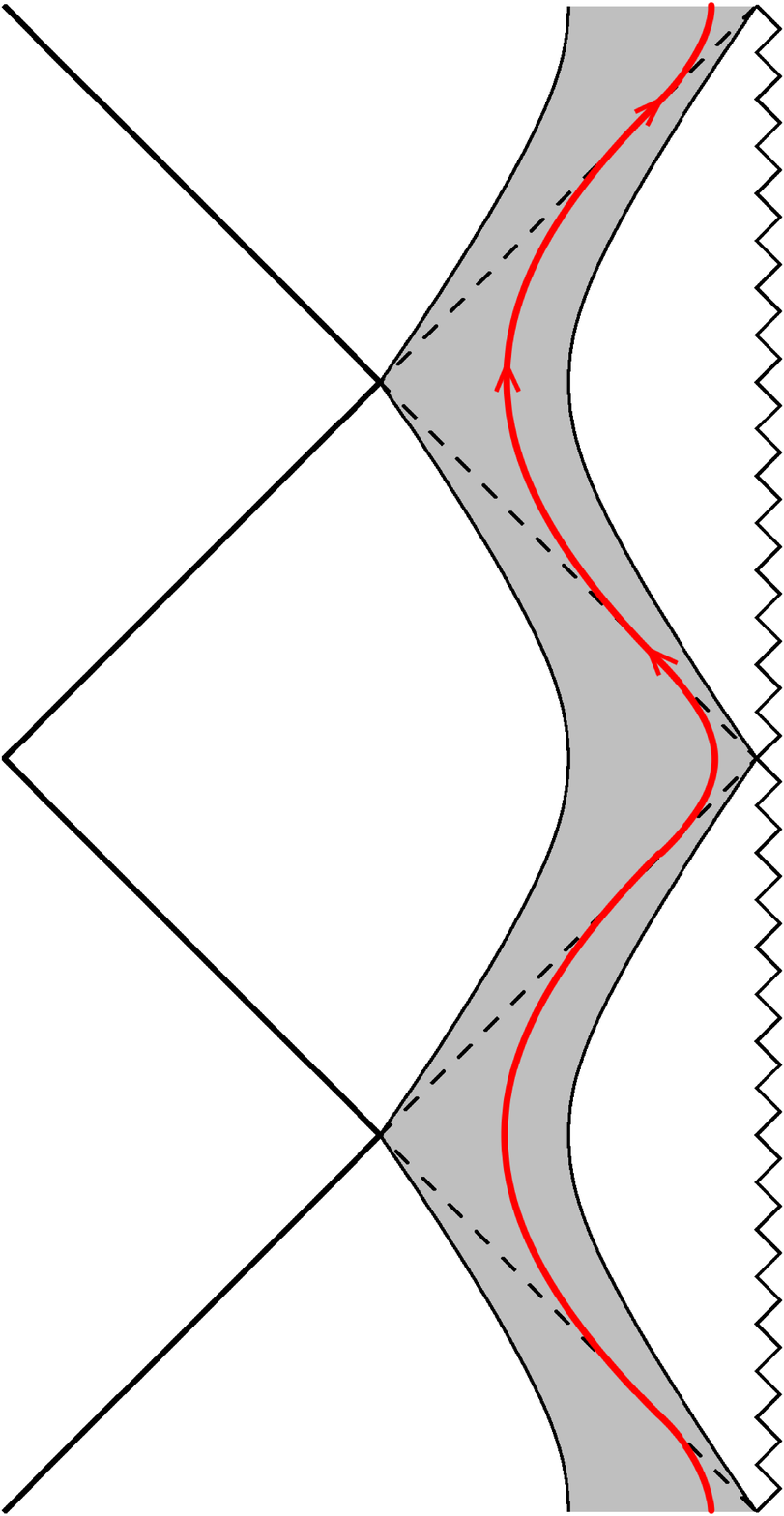}{2.0truein}

Supersymmetric chiral primary states were analyzed in detail in
\refs{\ssty,\gssy}, and come in several varieties.  Nonabelian
configurations of $N$ D0 branes can, via the Myers effect \myers,
correspond  to branes wrapping a non-trivial cycle $\Sigma$. In
defining a black hole ground state partition function, we wish to
fix the asymptotic charges, so we should not  sum over
configurations for which $\Sigma$ is a nontrivial cycle in $CY_3$.
This does not preclude configurations in which the $\Sigma$ is the
horizon $S^2$. Such a horizon-wrapped D2 does not lead to any
asymptotic D2 charge in the full black hole geometry, and so such
configurations should be summed over.  In fact we see below that
they are many of them at large $N$.

   The quantum mechanical
Hamiltonian and supercharges for various D0 configurations were
constructed in detail in \refs{\ssty,\gssy}. Here we summarize the
results for the case of interest, namely a D2 brane which wraps the
horizon $S^2$ and carries $N$ units of magnetic flux thereby
inducing $N$ units of D0 charge. Choosing the quantum mechanical
time to be Poincar\'e time, these see an $R\times CY_3$ target
space, with metric  \eqn\ruj{ds^2=T_{}\left({4Q d\xi^2 }+{\xi^2\over
Q}2g_{a \bar b}dz^adz^{\bar b}\right),} where $g_{a \bar b}$
describes the Ricci-flat attractor geometry \rts\ on $CY_3$, $T$ is
the mass of a wrapped D2 with $N$ units of flux
\eqn\jjf{T_{}={2\pi\over g_s}\sqrt{(4\pi Q^2)^2+N^2}} and $\xi$ is
the spatial coordinate in which the $AdS_2$ metric is
\eqn\adstwo{ds^2=Q^2{-dt^2+4\xi^2d\xi^2 \over \xi^4 }.}
 The D2
couples to the background RR flux $F^{(4)}$ of equation \rax. Since
this has two components tangent to the horizon, the D2 behaves like
a point particle in $CY_3$ with a two-form magnetic field
\eqn\rszk{F_{CY}=p^A \omega_A = {4\pi Q\over g_s}J.}

Altogether the bosonic part of the low-energy Hamiltonian
(generating $\p \over \p t$) for a single D2 is \eqn\rolk{H= {1
\over 8QT_{}}P_\xi^2 +{Q\over T_{}\xi^2}(P_a-A_a)g^{a \bar
b}(P_{\bar b}-A_{\bar b})+{32\pi^4Q^5\over g_s^2T\xi^2},}  where
$P_\xi, P_b$ are canonically conjugate to $\xi, z^b$.  There are no
angular fields on the $S^2$ in \rolk\ because the D2 wraps the
horizon. $A$ here is the gauge field on $CY_3$ obeying
\eqn\dij{dA=F_{CY}=p^A \omega_A.} The last term in \rolk\ is a
potential which repels the D2 from the $\xi=0$ $AdS_2$ boundary and
pushes it towards the black hole horizon at $\xi=\infty$. It is the
remnant of an imperfect cancellation between the brane mass and
coupling to RR gauge fields. The precise form given here is deduced
from supersymmetry in \gssy.

 The $SU(1,1|2)$ superisometries of the compactification must act
on the Hilbert space of the D2 quantum mechanics. Hence the
quantum mechanics has a complete set of superconformal generators
of (a central extension of) $SU(1,1|2)$ which are given in \gssy.
The generator of special conformal transformations is
\eqn\spec{K=2QT_{}{\xi^2}.} The generator of global $AdS_2$ time
translations is \eqn\rohg{L_0=H+K,} in which $K$ acts as a
potential barrier repelling the D2 from the black hole. Since $H$
pushes the D2 toward the black hole, $L_0$ eigenstates are
localized in the middle of the near-horizon $AdS_2$.

We are interested in the short multiplets of the superconformal
algebra. These multiplets have chiral primary states annihilated
by five of the eight superconformal charges, namely $G^{\alpha
\beta}_\half$ as well as \eqn\whx{\eqalign{ G^{++}_{-\half}&=
(QT_{})^{-\half}\left[{1\over 2}\lambda^{+}P_\xi - {i\over \xi}
{(L^\eta)^+}_\alpha \lambda^{\alpha} + {i\over 4\xi}
\bar\lambda^{+}\lambda_\alpha \lambda^\alpha + {i\over
4\xi}\lambda^{+}\right] \cr &~~~ +\left({Q\over T_{}}\right)^\half
\left[{\sqrt{2}\over \xi} (\eta^a)^{+} (P_a - A_a) - {8\pi^2
Q^2\over g_s} {i\over \xi}\lambda^+ \right]
 -(QT_{})^{\half} 2i\xi\lambda^{+} , }} where $\alpha,\beta=\pm$;
$\lambda^{\pm}, \bar\lambda^\pm$ are four worldvolume Goldstinos,
$(\eta^a)^{\pm}, (\bar\eta^{\bar a})^{\pm}$ are fermionic
collective coordinates on the $CY_3$ and $(L^\eta)_{\alpha\beta} =
g_{a\bar b}\eta^a_{(\alpha} \bar\eta^{\bar b}_{\beta)}$ are the
$SU(2)$ generators associated with the Lefschetz action on the
$CY_3$.

The chiral primaries have an $AdS_2$ and $CY_3$ component of their
wavefunction. We first consider the $CY_3$ component. The magnetic
field divides the $CY_3$ into $D={1 \over 6}\int F_{CY}\wedge
F_{CY}\wedge F_{CY}$ cells, corresponding the lowest Landau levels.
Put another way, the D2 sees a non-commutative Calabi-Yau with
non-commutativity parameter $\Theta^{-1}\sim F_{CY}$. There is
roughly one chiral primary for each cell. To be more precise, the
chiral primaries correspond to charged forms $h$ obeying
\eqn\cycond{ \bar D h = \bar D^* h =0 } where $iD_a \equiv P_a-A_a$.
Such forms are in one to one  correspondence with elements of
$H^q(CY_3, {\cal L}\otimes \Omega^p)$, where ${\cal L}$ is the line
bundle for which $c_1({\cal L})=[F_{CY}]$. For large $c_1$ and $q>0$
$H^q(CY_3, {\cal L}\otimes \Omega^p)$ vanishes. $\dim H^0(CY_3,
{\cal L}\otimes \Omega^p)$ can then be computed from the
Riemann-Roch formula, \eqn\dimcoh{ \eqalign{ h_0\equiv &\dim
H^0(CY_3, {\cal L}) = \int \bigl( {F_{CY}^3\over 6}+{c_2\wedge
F_{CY}\over 12} \bigr)\cr&~~~~~~~~= D + {1\over 12}c_2\cdot P,\cr
h_1\equiv & \dim H^0(CY_3, {\cal L}\otimes \Omega^1) = \int \bigl(
{F_{CY}^3\over 2}-{3c_2\wedge F_{CY}\over 4}+{c_3\over 2}
\bigr)\cr&~~~~~~~~= 3D -{3\over 4}c_2\cdot P-{\chi\over 2}, \cr
h_2\equiv & \dim H^0(CY_3, {\cal L}\otimes \Omega^2) = \int \bigl(
{F_{CY}^3\over 2}-{3c_2\wedge F_{CY}\over 4}-{c_3\over 2}
\bigr)\cr&~~~~~~~~~= 3D -{3\over 4}c_2\cdot P+{\chi\over 2}, \cr h_3
\equiv & \dim H^0(CY_3, {\cal L}\otimes \Omega^3) = \int \bigl(
{F_{CY}^3\over 6}+{c_2\wedge F_{CY}\over 12} \bigr)\cr&~~~~~~~~~= D
+ {1\over 12}c_2\cdot P. } } Note that the background flux $q_0$
does not appear in these formulae, so in counting chiral primaries
we can take $q_0\to 0$. \dimcoh\ may then be thought of as counting
chiral primaries carrying total electric D0 charge $N$ in the
attractor background produced by the magnetic D4 charges $p^A$.

The picture of Calabi-Yau cells here is reminiscent of the one
given in \refs{\juanbh,\fvijay}, in which BPS states are described
as D0 branes bound to the triple self-intersection sites of the D4
brane, except that here the explicit D4 branes are replaced by D4
fluxes. What has happened is that in the brane-geometry transition
the self-intersection sites have dissolved into lowest Landau
levels for the D0 branes or, equivalently, the cells of the
noncommutative Calabi-Yau. There may also ultimately be a relation
to the crystal atoms of \orv.

 Now we turn to the
$AdS_2$ component of the wavefunction. Given a form of the type
\cycond\ one can act on it with the Lefschetz raising operator
$(L^\eta)^{++}$ and obtain a new form $\tilde h$ that satisfies
\eqn\hwslfz{ D\tilde h = D^*\tilde h =0 } When the $CY_3$
component corresponds to a form of the type \hwslfz,
$G^{++}_{\pm\half}$ reduce to \eqn\rfji{G^{++}_{\pm\half} =
{\lambda^{++}\over \sqrt{QT}}\left[{1\over 2} P_\xi -{i\over \xi}
(L^\eta)^{+-} + {i\over 2\xi}\lambda^{+-}\lambda^{-+} + ({1\over
4}-{8\pi^2\over g_s} Q^3){i\over \xi}\mp 2iQT_{}\xi \right].}
These annihilate the state $|0\rangle$ defined by
$\lambda^{++}|0\rangle = 0$. When acting on $|0\rangle$,
$G^{--}_{\half}$ reduces to \eqn\gmmrd{ \eqalign{ G_{\half }^{--}
&= {\lambda^{--}\over \sqrt{QT}} \left[ {1\over 2}P_\xi + {i\over
\xi}(L^\eta)^{-+} - {i\over 2\xi} \lambda^{+-}\lambda^{-+} -
({1\over 4}-{8\pi^2\over g_s} Q^3){i\over \xi} - 2iQT_{}\xi
 \right] \cr &~~~~~- {i\over \xi \sqrt{QT}} (L^\eta)^{--}\lambda^{+-}. } } This is
solved by the $AdS_2$ wave function  \eqn\wsz{
|\psi\rangle=\xi^{2L^\eta_3 +16\pi^2  Q^3/g_s+\half}
\exp({-2QT_{}\xi^2})|0\rangle \otimes \tilde h } where the fermionic
part of the state obeys \eqn\wszl{ \lambda^{++}|0\rangle=
\lambda^{+-}|0\rangle=0.} \wsz\ is normalizable for positive Q. It
follows from the superconformal algebra \gssy\ that
$G^{+-}_{\half}|\psi\rangle=0$, and $|\chi\rangle = G^{-+}_\half
|\psi\rangle\not=0$. Now $|\chi\rangle$ is a chiral primary state.

We now count the multiparticle chiral primaries. First we consider
all possible ways that the D0s can form D2s.\foot{We ignore here the
possibility of mulitply-wrapped D2 bound states, which we have not
shown do not exist.} We partition the $N$ D0 branes into $k$
clusters of $n_i$ D0 branes each such that \eqn\rfs{\sum_{i=1}^k
n_i=N.} Each cluster then forms a wrapped D2 brane with $n_i$ units
of magnetic flux. Each of the $k$ such D2 branes can then sit in one
of the $h_0+h_1+h_2+h_3$ chiral primary states. Noting the chiral
primary states associated to $h_p$ have quantum mechanical fermion
number $p$, the counting of such configurations is the same as the
counting of states of a 1+1 CFT with $h_0+h_2$ bosons, $h_1+h_3$
fermions and total left-moving momentum $N$. The chiral primary
generating  function is then \eqn\acw{Z={\rm Tr}q^N = \prod_n
{(1+q^n)^{h_1+h_3} \over (1-q^n)^{h_0+h_2}}, } where the trace is
over chiral primaries. Using the values \dimcoh\ for $h_r$ and the
well known asymptotics of \acw\ we find an asymptotic formula for
entropy as a function of $N$ \eqn\poy{S=2\pi \sqrt{D_{ABC}p^Ap^Bp^C
N}, }  in precise agreement with the macroscopic area law for a
black hole with D4 charge $p^A$ and D0 charge $N$. This is evidence
for the proposal that near horizon chiral primaries count black hole
ground states.

Let us now briefly comment on what would be needed to promote this
to a full-blown $AdS_2/CFT_1$ correspondence. First there are
various ${1 \over N}$ corrections to the BPS degeneracies. These
come for example from the subleading terms in \dimcoh\ as well as
configurations with D0 charge but no wrapped D2s.  The
superconformal quantum mechanics also has computable non-BPS
excitations. This might be identified with near-extremal black hole
microstates. The total D0 charge of the system is $N+q_0$, the sum
of the number of explicit D0 branes plus the background flux. In the
non-BPS sector the quantum mechanics has $q_0$ dependence, so one
must understand the $q_0\to 0$ limit in which the target $CY_3$
scales to zero size, among other issues.

\centerline{\bf Acknowledgements} This work was supported in part
by DOE grant DE-FG02-91ER40654. We are grateful to A. Simons for
useful conversations.

\listrefs

\end